\documentclass{pasj01}

\def\commenta{$^*$}
\def\commentb{$^\dagger$}

\newcounter{author}
\def\authorcount#1#2{\refstepcounter{author}\label{#1}
                     \altaffiltext{\ref{#1}}{#2}}

\begin{document}
\SetRunningHead{T. Kato et al.}{RZ Leonis Minoris}

\Received{201X/XX/XX}
\Accepted{201X/XX/XX}

\title{RZ Leonis Minoris Bridging between ER Ursae Majoris-Type
       Dwarf Nova and Novalike System}

\author{Taichi~\textsc{Kato},\altaffilmark{\ref{affil:Kyoto}*}
        Ryoko~\textsc{Ishioka},\altaffilmark{\ref{affil:ASIAA}}
        Keisuke~\textsc{Isogai},\altaffilmark{\ref{affil:Kyoto}}
        Mariko~\textsc{Kimura},\altaffilmark{\ref{affil:Kyoto}}
        Akira~\textsc{Imada},\altaffilmark{\ref{affil:HidaKwasan}}
        Ian~\textsc{Miller},\altaffilmark{\ref{affil:Miller}}
        Kazunari~\textsc{Masumoto},\altaffilmark{\ref{affil:OKU}}
        Hirochika~\textsc{Nishino},\altaffilmark{\ref{affil:OKU}}
        Naoto~\textsc{Kojiguchi},\altaffilmark{\ref{affil:OKU}}
        Miho~\textsc{Kawabata},\altaffilmark{\ref{affil:OKU}}
        Daisuke~\textsc{Sakai},\altaffilmark{\ref{affil:OKU}}
        Yuki~\textsc{Sugiura},\altaffilmark{\ref{affil:OKU}}
        Hisami~\textsc{Furukawa},\altaffilmark{\ref{affil:OKU}}
        Kenta~\textsc{Yamamura},\altaffilmark{\ref{affil:OKU}}
        Hiroshi~\textsc{Kobayashi},\altaffilmark{\ref{affil:OKU}}
        Katsura~\textsc{Matsumoto},\altaffilmark{\ref{affil:OKU}}
        Shiang-Yu~\textsc{Wang},\altaffilmark{\ref{affil:ASIAA}}
        Yi~\textsc{Chou},\altaffilmark{\ref{affil:NCU}}
        Chow-Choong~\textsc{Ngeow},\altaffilmark{\ref{affil:NCU}}
        Wen-Ping~\textsc{Chen},\altaffilmark{\ref{affil:NCU}}
        Neelam~\textsc{Panwar},\altaffilmark{\ref{affil:NCU}}
        Chi-Sheng~\textsc{Lin},\altaffilmark{\ref{affil:NCU}}
        Hsiang-Yao~\textsc{Hsiao},\altaffilmark{\ref{affil:NCU}}
        Jhen-Kuei~\textsc{Guo},\altaffilmark{\ref{affil:NCU}}
        Chien-Cheng~\textsc{Lin},\altaffilmark{\ref{affil:NCU}}
        Chingis~\textsc{Omarov},\altaffilmark{\ref{affil:TSHAO1}}
        Anatoly~\textsc{Kusakin},\altaffilmark{\ref{affil:TSHAO1}}
        Maxim~\textsc{Krugov},\altaffilmark{\ref{affil:TSHAO1}}
        Donn~R.~\textsc{Starkey},\altaffilmark{\ref{affil:Starkey}}
        Elena~P.~\textsc{Pavlenko},\altaffilmark{\ref{affil:CrAO}}
        Kirill~A.~\textsc{Antonyuk},\altaffilmark{\ref{affil:CrAO}}
        Aleksei~A.~\textsc{Sosnjvskij},\altaffilmark{\ref{affil:CrAO}}
        Oksana~I.~\textsc{Antonyuk},\altaffilmark{\ref{affil:CrAO}}
        Nikolai~V.~\textsc{Pit},\altaffilmark{\ref{affil:CrAO}}
        Alex~V.~\textsc{Baklanov},\altaffilmark{\ref{affil:CrAO}}
        Julia~V.~\textsc{Babina},\altaffilmark{\ref{affil:CrAO}}
        Hiroshi~\textsc{Itoh},\altaffilmark{\ref{affil:Ioh}}
        Stefano~\textsc{Padovan},\altaffilmark{\ref{affil:AAVSO}}
        Hidehiko~\textsc{Akazawa},\altaffilmark{\ref{affil:OUS}}
        Stella~\textsc{Kafka},\altaffilmark{\ref{affil:AAVSO}}
        Enrique~de~\textsc{Miguel},\altaffilmark{\ref{affil:Miguel}}$^,$\altaffilmark{\ref{affil:Miguel2}}
        Roger~D.~\textsc{Pickard},\altaffilmark{\ref{affil:BAAVSS}}$^,$\altaffilmark{\ref{affil:Pickard}}
        Seiichiro~\textsc{Kiyota},\altaffilmark{\ref{affil:Kis}}
        Sergey~Yu.~\textsc{Shugarov},\altaffilmark{\ref{affil:Sternberg}}$^,$\altaffilmark{\ref{affil:Slovak}}
        Drahomir~\textsc{Chochol},\altaffilmark{\ref{affil:Slovak}}
        Viktoriia~\textsc{Krushevska},\altaffilmark{\ref{affil:MainUkraine}}
        Matej~\textsc{Seker\'a\v{s}},\altaffilmark{\ref{affil:Slovak}}
        Olga~\textsc{Pikalova},\altaffilmark{\ref{affil:MSUPhys}}
        Richard~\textsc{Sabo},\altaffilmark{\ref{affil:Sabo}}
        Pavol~A.~\textsc{Dubovsky},\altaffilmark{\ref{affil:Dubovsky}}
        Igor~\textsc{Kudzej},\altaffilmark{\ref{affil:Dubovsky}}
        Joseph~\textsc{Ulowetz},\altaffilmark{\ref{affil:Ulowetz}}
        Shawn~\textsc{Dvorak},\altaffilmark{\ref{affil:Dvorak}}
        Geoff~\textsc{Stone},\altaffilmark{\ref{affil:AAVSO}}
        Tam\'as~\textsc{Tordai},\altaffilmark{\ref{affil:Polaris}}
        Franky~\textsc{Dubois},\altaffilmark{\ref{affil:Dubois}}
        Ludwig~\textsc{Logie},\altaffilmark{\ref{affil:Dubois}}
        Steve~\textsc{Rau},\altaffilmark{\ref{affil:Dubois}}
        Siegfried~\textsc{Vanaverbeke},\altaffilmark{\ref{affil:Dubois}}
        Tonny~\textsc{Vanmunster},\altaffilmark{\ref{affil:Vanmunster}}
        Arto~\textsc{Oksanen},\altaffilmark{\ref{affil:Nyrola}}
        Yutaka~\textsc{Maeda},\altaffilmark{\ref{affil:Mdy}}
        Kiyoshi~\textsc{Kasai},\altaffilmark{\ref{affil:Kai}}
        Natalia~\textsc{Katysheva},\altaffilmark{\ref{affil:Sternberg}}
        Etienne~\textsc{Morelle},\altaffilmark{\ref{affil:Morelle}}
        Vitaly~V.~\textsc{Neustroev},\altaffilmark{\ref{affil:Neustroev1}}$^,$\altaffilmark{\ref{affil:Neustroev2}}
        George~\textsc{Sjoberg},\altaffilmark{\ref{affil:Sjoberg}}$^,$\altaffilmark{\ref{affil:AAVSO}}
}

\authorcount{affil:Kyoto}{
     Department of Astronomy, Kyoto University, Kyoto 606-8502, Japan}
\email{$^*$tkato@kusastro.kyoto-u.ac.jp}

\authorcount{affil:ASIAA}{
     Institute of Astronomy and Astrophysics, Academia Sinica,
     11F of Astronomy-Mathematics Building, National Taiwan University,
     No. 1, Sec. 4, Roosevelt Rd, Taipei 10617, Taiwan
}

\authorcount{affil:HidaKwasan}{
     Kwasan and Hida Observatories, Kyoto University, Yamashina,
     Kyoto 607-8471, Japan}

\authorcount{affil:Miller}{
     Furzehill House, Ilston, Swansea, SA2 7LE, UK}

\authorcount{affil:OKU}{
     Osaka Kyoiku University, 4-698-1 Asahigaoka, Osaka 582-8582, Japan}

\authorcount{affil:NCU}{
     Graduate Institute of Astronomy, National Central University,
     Jhongli 32001, Taiwan}

\authorcount{affil:TSHAO1}{
     Fessenkov Astrophysical Institute, Observatory 23, Almaty, 050020,
     Kazakhstan}

\authorcount{affil:Starkey}{
     DeKalb Observatory, H63, 2507 County Road 60, Auburn, Indiana 46706, USA}

\authorcount{affil:CrAO}{
     Federal State Budget Scientific Institution ``Crimean Astrophysical
     Observatory of RAS'', Nauchny, 298409, Republic of Crimea}

\authorcount{affil:Ioh}{
     Variable Star Observers League in Japan (VSOLJ),
     1001-105 Nishiterakata, Hachioji, Tokyo 192-0153, Japan}

\authorcount{affil:OUS}{
     Department of Biosphere-Geosphere System Science, Faculty of Informatics,
     Okayama University of Science, 1-1 Ridai-cho, Okayama,
     Okayama 700-0005, Japan}

\authorcount{affil:AAVSO}{
     American Association of Variable Star Observers, 49 Bay State Rd.,
     Cambridge, MA 02138, USA}

\authorcount{affil:Miguel}{
     Departamento de Ciencias Integradas, Facultad de Ciencias
     Experimentales, Universidad de Huelva,
     21071 Huelva, Spain}

\authorcount{affil:Miguel2}{
     Center for Backyard Astrophysics, Observatorio del CIECEM,
     Parque Dunar, Matalasca\~nas, 21760 Almonte, Huelva, Spain}

\authorcount{affil:BAAVSS}{
     The British Astronomical Association, Variable Star Section (BAA VSS),
     Burlington House, Piccadilly, London, W1J 0DU, UK}

\authorcount{affil:Pickard}{
     3 The Birches, Shobdon, Leominster, Herefordshire, HR6 9NG, UK}

\authorcount{affil:Kis}{
     VSOLJ, 7-1 Kitahatsutomi, Kamagaya, Chiba 273-0126, Japan}

\authorcount{affil:Sternberg}{
     Sternberg Astronomical Institute, Lomonosov Moscow State University, 
     Universitetsky Ave., 13, Moscow 119992, Russia}

\authorcount{affil:Slovak}{
     Astronomical Institute of the Slovak Academy of Sciences,
     05960 Tatranska Lomnica, Slovakia}

\authorcount{affil:MainUkraine}{
     Main astronomical observatory of the National Academy of Sciences of
     Ukraine, 27 Akademika Zabolotnoho ave., 03680 Kyiv, Ukraine}

\authorcount{affil:MSUPhys}{
     Faculty of Physics, Lomonosov Moscow
     State University, Leninskie Gory, Moscow, 119991, Russia}

\authorcount{affil:Sabo}{
     2336 Trailcrest Dr., Bozeman, Montana 59718, USA}

\authorcount{affil:Dubovsky}{
     Vihorlat Observatory, Mierova 4, 06601 Humenne, Slovakia}

\authorcount{affil:Ulowetz}{
     Center for Backyard Astrophysics Illinois,
     Northbrook Meadow Observatory, 855 Fair Ln, Northbrook,
     Illinois 60062, USA}

\authorcount{affil:Dvorak}{
     Rolling Hills Observatory, 1643 Nightfall Drive,
     Clermont, Florida 34711, USA}

\authorcount{affil:Polaris}{
     Polaris Observatory, Hungarian Astronomical Association,
     Laborc utca 2/c, 1037 Budapest, Hungary}

\authorcount{affil:Dubois}{
     Public observatory Astrolab Iris, Verbrandemolenstraat 5,
     B 8901 Zillebeke, Belgium}

\authorcount{affil:Vanmunster}{
     Center for Backyard Astrophysics Belgium, Walhostraat 1A,
     B-3401 Landen, Belgium}

\authorcount{affil:Nyrola}{
     Hankasalmi observatory, Vertaalantie 116, FIN-40500 Hankasalmi,
     Finland}

\authorcount{affil:Mdy}{
     Kaminishiyamamachi 12-14, Nagasaki, Nagasaki 850-0006, Japan}

\authorcount{affil:Kai}{
     Baselstrasse 133D, CH-4132 Muttenz, Switzerland}

\authorcount{affil:Morelle}{
     9 rue Vasco de GAMA, 59553 Lauwin Planque, France}

\authorcount{affil:Neustroev1}{
     Finnish Centre for Astronomy with ESO (FINCA), University of
     Turku, V\"{a}is\"{a}l\"{a}ntie 20, FIN-21500 Piikki\"{o}, Finland}

\authorcount{affil:Neustroev2}{
     Astronomy research unit, PO Box 3000, FIN-90014 University of
     Oulu, Finland}

\authorcount{affil:Sjoberg}{
     The George-Elma Observatory, 9 Contentment Crest, \#182,
     Mayhill, New Mexico 88339, USA}



\maketitle

\begin{abstract}
We observed RZ LMi, which is renowned for the extremely
($\sim$19~d) short supercycle and is a member of
a small, unusual class of cataclysmic variables
called ER UMa-type dwarf novae, in 2013 and 2016.
In 2016, the supercycles of this object substantially
lengthened in comparison to the previous measurements
to 35, 32, 60~d for three consecutive
superoutbursts.  We consider that the object virtually
experienced a transition to the novalike state
(permanent superhumper).
This observed behavior extremely well reproduced
the prediction of the thermal-tidal instability model.
We detected a precursor in the 2016 superoutburst
and detected growing (stage A) superhumps with
a mean period of 0.0602(1)~d  in 2016
and in 2013.  Combined with the period of superhumps
immediately after the superoutburst, the mass ratio is not
as small as in WZ Sge-type dwarf novae, having orbital
periods similar to RZ LMi.
By using least absolute shrinkage and selection
operator (Lasso) two-dimensional power spectra,
we detected possible negative superhumps with
a period of 0.05710(1)~d.  We estimated
the orbital period of 0.05792~d, which
suggests a mass ratio of 0.105(5).  This relatively
large mass ratio is even above ordinary SU UMa-type
dwarf novae, and it is also possible that
the exceptionally high mass-transfer rate in RZ LMi
may be a result of a stripped core evolved secondary
which are evolving toward an AM CVn-type object.
\end{abstract}

\section{Introduction}

   SU UMa-type dwarf novae (DNe) are a class of cataclysmic
variables (CVs) which are close binary systems
transferring matter from a red dwarf secondary to
a white dwarf, forming an accretion disk.
In SU UMa-type dwarf novae, two types of outbursts
are seen: normal outbursts and superoutbursts.
Superoutbursts are defined by the presence of
(positive) superhumps, which are humps with a period a few percent
longer than the orbital period
[for general information of CVs, DNe, SU UMa-type 
dwarf novae and superhumps, see e.g. \citet{war95book}].

   RZ LMi is one of the most enigmatic SU UMa-type dwarf novae.
This object was originally discovered as an ultraviolet-excess
variable star (\cite{lip81FBSCV};
later given a designation of FBS 0948$+$344: \cite{abr93FBS}).
\citet{lip81FBSCV} reported that the object had a spectral
energy distribution of spectral classes O--B at maximum.
When the object was fading or rising, strong emission lines
appeared.  \citet{lip81FBSCV} studied 16 objective prism plates
and seven direct imaging plates and recorded strong variation
with a range of 14--17 mag.  The object was recorded at
17 mag on the Palomar Obervatory Sky Survey (POSS).
Based on the colors and rapid variation within
a few days, \citet{lip81FBSCV} suggested 
a dwarf nova-type classification.
\citet{gre82PGsurveyCV} also selected it as an ultraviolet-excess
object (PG 0948$+$344) and was spectroscopically confirmed
as a CV.  \citet{gre82PGsurveyCV} gave a variability range of
$B$=14.4--16.8 without details (the minimum likely referred
to the magnitude in POSS).  Despite the finding by
\citet{lip81FBSCV}, \citet{NameList67} gave a designation
of RZ LMi as a novalike (NL) variable by referring to
\citet{gre82PGsurveyCV}.  \citet{kon84KUV2} also selected
RZ LMi as an ultraviolet-excess object.

   The nature of the variability of this object remained
unclear.  Knowing the dwarf nova-type classification,
one of the authors (TK) visually observed this object
in 1987--1988 and found relatively stable $\sim$20~d
cycle lengths but probably with additional short outbursts.
This result was reported in the domestic variable star
bulletin ``Henkousei'' (in Japanese).  This report was
probably the first to document the unusual cyclic variation
of RZ LMi.  Two teams photometrically observed this object
and \citet{rob94rzlmi} reported an unusual long-term
repetitive light curve with a stable period (it is
now known as the supercycle) of 19.2~d.
\citet{pik95rzlmi} reported that the behavior was different
from a typical dwarf nova in that it showed short
fading and brightening.  \citet{pik95rzlmi} reported
a possible period of 21.167~d or 23.313~d and classified
it to be a VY Scl-type NL.

   A clue to understanding the unusual behavior of RZ LMi
came from the discovery of the SU UMa-type dwarf nova
ER UMa with an ultrashort supercycle (interval between
superoutbursts) of 43~d in 1994 \citep{kat95eruma}.
As orally presented in the conference held in Abano Terme,
Italy, 20--24 June 1994, Robertson's team originally
considered that the recurring NL-type bright state and 
the dwarf nova-type state in ER UMa and RZ LMi was
the best example to show the consequences of
recurring mass-transfer bursts from the secondary.
In this conference after Robertson's presentation,
however, Y. Osaki introduced Kato and Kunjaya's detection of
superhumps in ER UMa and showed that the unusual behavior of
ER UMa could be understood within the framework
of the thermal-tidal instability (TTI) model \citep{osa89suuma}
if one can allow an exceptionally high mass-transfer rate
(cf. \cite{osa95eruma}).
After this conference, Robertson's team published
a paper following our interpretation
(\cite{rob94rzlmi}; \cite{rob95eruma}).\footnote{
   The dwarf nova-type nature was first clarified by
   \citet{iid94eruma}.
   \citet{mis95PGCV} also observed ER UMa in 1993--1994 and
   finally reached the same conclusion as \citet{kat95eruma}.
}
After the discovery of superhumps in ER UMa and
the recognition of the similarity between ER UMa and RZ LMi,
superhumps were naturally sought --- the competition was
intense: both \citet{rob95eruma} and \citet{nog95rzlmi}
observed the same superoutburst in 1995 and detected
superhumps.  These observations confirmed that RZ LMi
does indeed belongs to the SU UMa-type dwarf novae.
These objects, together with V1159 Ori, are usually
called ER UMa-type stars (cf. \cite{kat99erumareview}).

   The mechanism of the ultrashort (19~d) supercycle
and the unusually regular outburst pattern remained
a mystery.  \citet{rob94rzlmi} suspected a mechanism
outside the disk in addition to the one in the disk.
A later publication by \citet{ole08rzlmi} followed
the former possibility and suggested a third body.
In the standard TTI model, this short supercycle is
difficult to reproduce, and \citet{osa95rzlmi} presented
a working hypothesis that in RZ LMi the disk becomes
thermally unstable during the superoutburst earlier
than in other SU UMa-type dwarf novae.
\citet{hel01eruma} suggested, following the interpretation
in \citet{osa95rzlmi}, that in systems with very small
mass ratios ($q$), the tidal torque is too small
to maintain the superoutburst, and that there occurs
a decoupling between the thermal and tidal instabilities.
\citet{hel01eruma} suggested that one of the consequences
of this decoupling can be found as the persistent
superhumps after superoutbursts.

   According to the working hypotheses by \citet{osa95rzlmi}
and \citet{hel01eruma}, it is strongly predicted that
the $q$ is small in RZ LMi and that the disk radius after
the superoutburst is larger than in other SU UMa-type
dwarf novae.  RZ LMi, however, has defied every attempt
to determine the orbital period, since it mostly stays
in the ``outburst'' state and it is difficult to make
a radial-velocity study in short quiescence.
The almost continuous presence of superhumps also made it
difficult to detect potential orbital variations by
photometric methods.
Without the orbital period or a radial-velocity study,
it remained impossible to observationally determine $q$ and
the evolutionary status of RZ LMi remained unclear
despite its unusual outburst properties.

   The situation dramatically changed after
the detection of a possible change in the outburst
pattern in the AAVSO observations (vsnet-alert 19524).
We have conducted a world-wide campaign to observe RZ LMi
during the 2016 season.  The new development
has also been helped by the classification of superhumps
stage (A, B and C: \cite{Pdot}) and identification
of stage A superhumps representing the growing phase
of superhumps at the radius of the 3:1 resonance
(\cite{osa13v344lyrv1504cyg}; \cite{kat13qfromstageA}).

   In this paper, key information and results are given
in the main paper.  The results not directly
related to the conclusion of the paper, such as
variation of superhump periods and variation of
the profile of superhumps, are given in
the Supplementary Information since
they would provide useful information to
expert readers.

\section{Observations}\label{sec:obs}

   The data were obtained under campaigns led by 
the VSNET Collaboration \citep{VSNET} in 2016.
We also used the public data from 
the AAVSO International Database\footnote{
   $<$http://www.aavso.org/data-download$>$.
}.
Time-resolved photometric observations were obtained
on 75 nights (nights with more than 100 observations)
between February 25 and June 10.  Some snapshot observations
were obtained on some other nights.
We also performed time-resolved photometric observations during
the period of 2013 March 5--April 29 in 2013 at 9 sites.
We also obtained snapshot observations (several observations
typically $\sim$20 min apart) on 47 nights
between 2014 March 8 and May 22.
The logs of observations are given in e-tables.
Alphabetical abbreviated codes in the log are
observer codes of AAVSO and they mean that the data
were taken from the AAVSO database.
We also used historical photographic data
reported by \citet{pik95rzlmi} (listed as ``Shugarov'',
who finally compiled all the data, in this paper).

   The data were analyzed just in the same way as described
in \citet{Pdot} and \citet{Pdot6}.  We mainly used
R software\footnote{
   The R Foundation for Statistical Computing:\\
   $<$http://cran.r-project.org/$>$.
} for data analysis.
In de-trending the data, we divided the data into
four segments in relation to the outburst phase and
used locally-weighted polynomial regression 
(LOWESS: \cite{LOWESS}).
The times of superhumps maxima were determined by
the template fitting method as described in \citet{Pdot}.
The times of all observations are expressed in 
barycentric Julian Days (BJD).

\begin{figure*}
  \begin{center}
    \FigureFile(110mm,170mm){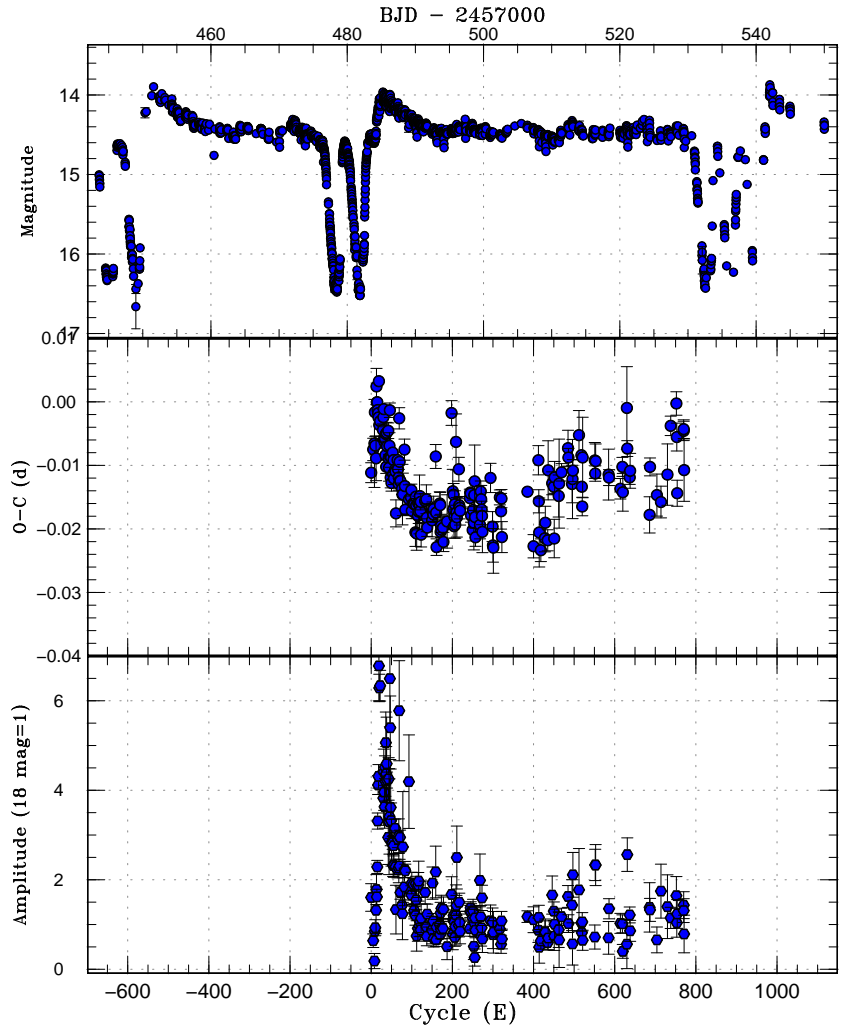}
  \end{center}
  \caption{Light curve and $O-C$ diagram of superhumps in RZ LMi (2016).
     (Upper:) Light curve.  The data were binned to 0.0023~d.
     In this figure, we show three consecutive superoutbursts
     we observed.
     (Middle:) $O-C$ diagram for the second superoutburst
     (filled circles).
     We used a period of 0.05955~d for calculating the $O-C$ residuals.
     (Lower:) Amplitudes of superhumps.  The scale is linear
     and the pulsed flux is shown in a unit corresponding
     to 18 mag = 1.
  }
  \label{fig:rzlmihumpall2}
\end{figure*}

\section{Results and Discussion}\label{sec:results}

\subsection{Outburst light curve and emergence of superhumps}\label{sec:outburst}

   Figure \ref{fig:rzlmihumpall2} illustrates the
three consecutive superoutbursts in 2016 
during which we obtained time-resolved
photometric observations.  The initial part
of the light curve (upper panel) was the final fading
part of the preceding superoutburst.  There was only one
normal outburst between the first and second superoutbursts.
Although the initial part of the first superoutburst
was not well sampled, we found the superoutburst lasted for 26~d.

   There were two normal outbursts between the second
and the third superoutburst.  The duration of the second
superoutburst was 48~d, even longer than the first one.
The intervals (supercycles) between the three superoutbursts
were 32~d and 60~d.  These values were 2--3 times longer
than the historical supercycle (19~d) of this object
(\cite{rob94rzlmi}; \cite{nog95rzlmi}).

\begin{figure*}
  \begin{center}
    \FigureFile(110mm,170mm){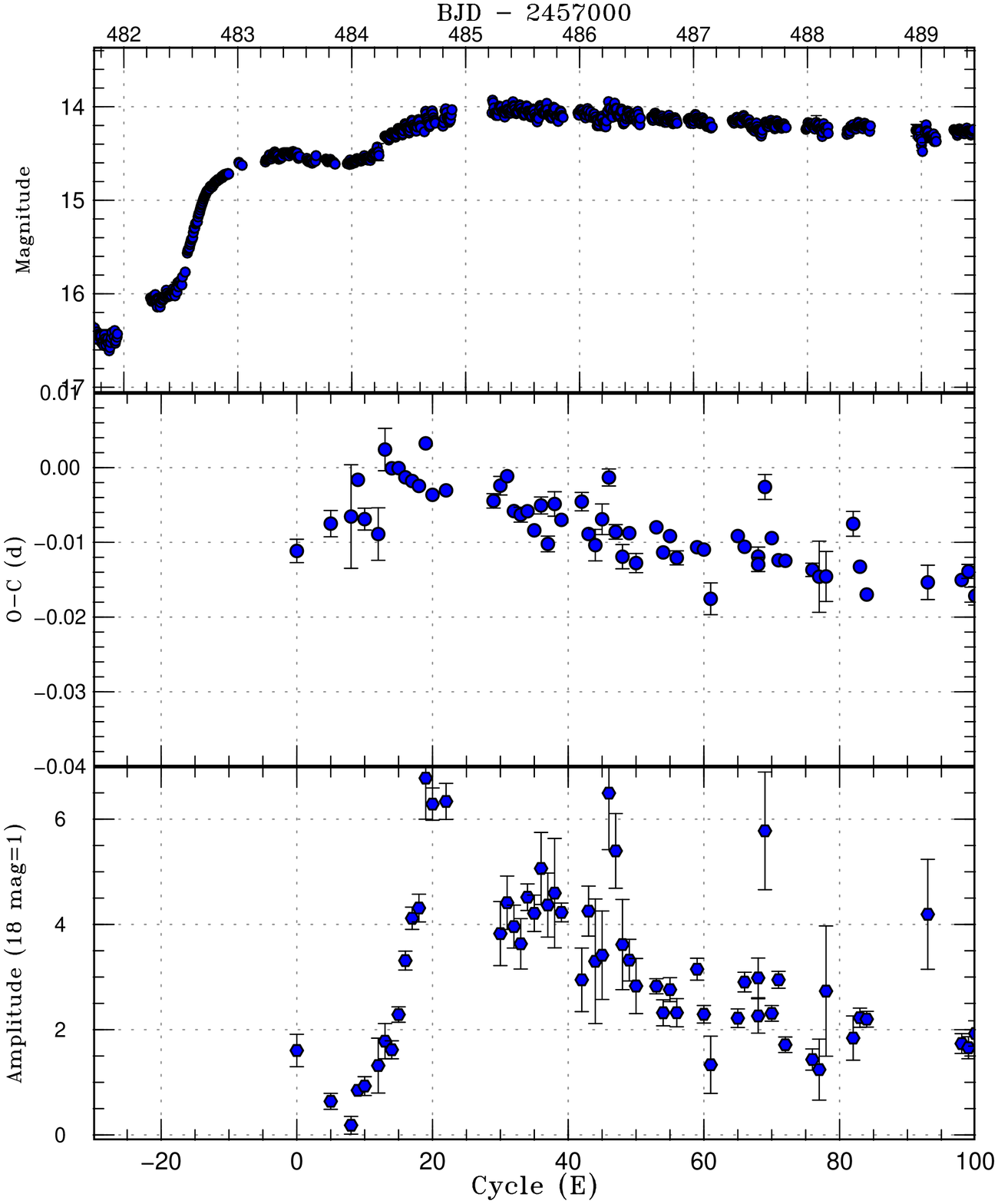}
  \end{center}
  \caption{Light curve and $O-C$ diagram of superhumps in RZ LMi (2016).
     This figure is an enlargement of figure \ref{fig:rzlmihumpall2}.
     (Upper:) Light curve.  The data were binned to 0.0023~d.
     The precursor part is clearly seen between BJD 2457483
     and 2457484.  During this phase, superhumps rapidly grew.
     (Middle:) $O-C$ diagram for the second superoutburst
     (filled circles).
     We used a period of 0.05955~d for calculating the $O-C$ residuals.
     (Lower:) Amplitudes of superhumps.  The scale is linear
     and the pulsed flux is shown in a unit corresponding
     to 18 mag = 1.
  }
  \label{fig:rzlmihumpall2a}
\end{figure*}

\begin{figure*}
  \begin{center}
    \FigureFile(130mm,80mm){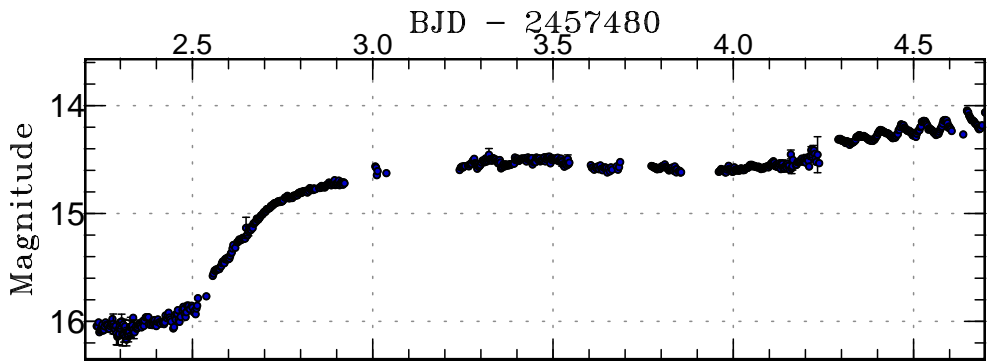}
  \end{center}
  \caption{Growing superhumps and precursor
  at the start of the 2016 April superoutburst.
  The data were binned to 0.0023~d.
  Superhump grew between BJD 2457483 and 2457484.
  }
  \label{fig:rzlmi2016rise}
\end{figure*}

   During the best observed second superoutburst,
there was a shoulder (precursor outburst).
Figure \ref{fig:rzlmihumpall2a} shows the initial
part of this superoutburst.  The precursor part is
clearly seen between BJD 2457483 and 2457484.
During this phase, superhumps rapidly grew
to the maximum amplitude and they started to decay
slowly (lower panell; see also figure \ref{fig:rzlmi2016rise}
for an enlargement).  This behavior is the same
as in ordinary SU UMa-type dwarf novae [see e.g.
figure 19 in \citet{Pdot4} for one of the best known
SU UMa-type dwarf novae VW Hyi; see also figure 4
in \citet{osa13v1504cygKepler} for the corresponding
part of the Kepler data of V1504 Cyg].
Similar precursors were also recorded in snapshot
observations of the first superoutburst and less
markedly (fewer observations) during
the third superoutburst.  These observations indicate
that RZ LMi shows precursor--main superoutburst
as in ordinary SU UMa-type dwarf novae at least
in this season.

   The variable supercycles now safely excludes
the previously supposed stable clock
(\cite{rob94rzlmi}; \cite{ole08rzlmi}) to produce
regular superoutbursts.  The present finding completely
excludes the third body for producing such stable
periodicity (cf. \cite{ole08rzlmi}).
Furthermore, \citet{osa95eruma} already showed that
the supercycle length once reaches the minimum
as the mass-transfer rate ($\dot{M}$) increases,
but that it lengthens again as $\dot{M}$ further
increases.  As $\dot{M}$ increases, the system
eventually reaches the ``permanent outburst''
state (see figure 2 in \cite{osa95eruma}), bridging
ER UMa-type objects and what is called
permanent superhumpers (\cite{ski93bklyn}; \cite{pat99SH}).
The current state of RZ LMi exactly reproduced
this prediction, and it provides a strong support
to explain the unusual short supercycles in
ER UMa-type objects.

   \citet{osa95rzlmi} formulated the duration
of a superoutburst ($t_{\rm supermax}$) as below:
\begin{equation}
t_{\rm supermax} \sim t_{\rm vis}[f_M/(1-1/e)][1-(\dot{M}/\dot{M}_{\rm crit})]^{-1/2},
\label{equ:tsuper}
\end{equation}
where $t_{\rm vis}$ is the viscous depletion timescale
and $\dot{M}_{\rm crit}$ is the critical $\dot{M}$
to produce a hot, stable disk, respectively.
The factor $f_M$ is the fraction of the disk mass
accreted during a superoutburst.  It is given by
\begin{equation}
f_M \simeq 1-(R_0/R_{\rm d,crit})^{3.0},
\label{equ:fracmass}
\end{equation}
where $R_0$ and $R_{\rm d,crit}$ represent the disk
radius at the end of a superoutburst and at the start
of a superoutburst (assuming that the disk critically
reaches the radius of the 3:1 resonance at the start
of a superoutburst), respectively.
If we consider that $t_{\rm vis}$ and $R_0$ are the same
between different superoutbursts of RZ LMi,
we can estimate $\dot{M}$ during the current state.
Using the parameters in \citet{osa95rzlmi},
$t_{\rm vis}$=11.2~d and an assumption of a large disk
radius at the end of a superoutburst $R_0$=0.42$a$,
where $a$ is the binary separation, the historical
$t_{\rm supermax}$ of 6~d \citep{rob95eruma} is reproduced with
$\dot{M}/\dot{M}_{\rm crit}$=0.5.
The current $t_{\rm supermax}$ values of 26~d and 48~d
require $\dot{M}/\dot{M}_{\rm crit}$=0.97 and 0.99,
respectively.  Although these estimates have uncertainties
due to various assumptions, it is certain that the current
state of RZ LMi is {\it critically} close to
the stability border.  If RZ LMi increases $\dot{M}$
further by 1\%, the object should become a permanent
superhumper.  As judged from these estimates, we have seen
an almost complete transition from an ER UMa-type object
to a permanent superhumper.  There has been at least
one case (BK Lyn) in which a permanent superhumper
became an ER UMa-type object and then returned back
(\cite{pat13bklyn}; \cite{Pdot4}) and the case of
RZ LMi may not be special.

\subsection{Growing (stage A) superhumps and post-superoutburst superhumps}\label{sec:stageA}

\begin{figure}
  \begin{center}
    \FigureFile(80mm,110mm){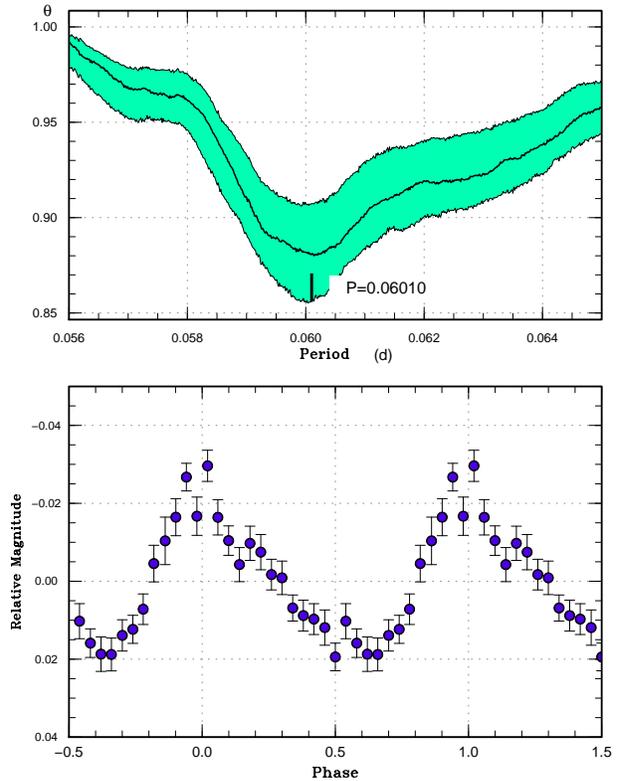}
  \end{center}
  \caption{Stage A superhumps in RZ LMi (2016).
     The data between BJD 2457483.01 and 2457484.40
     were used.
     (Upper): PDM analysis.  We analyzed 100 samples which
     randomly contain 50\% of observations, and performed PDM
     analysis for these samples.
     The bootstrap result is shown as a form of 90\% confidence
     intervals in the resultant PDM $\theta$ statistics.
     (Lower): Phase-averaged profile.}
  \label{fig:rzlmi2016shapdm}
\end{figure}

   As is best seen in the lower panel of
figure \ref{fig:rzlmihumpall2a},
the amplitudes of superhumps rapidly grew in $\sim$20
cycles.  During the initial $\sim$13 cycles, the $O-C$
values (middle panel) were negative and a characteristic
kink around $E$=13 indicates that this object showed
long-period (stage A) superhumps before entering
the stable phase of stage B superhumps with a shorter
period as in other SU UMa-type dwarf novae
(see e-tables
for the full list of times of superhump maxima).
This identification appears particularly confident
since it has been recently demonstrated that ER UMa,
the prototype of ER UMa-type objects, showed the same pattern
with a precursor outburst \citep{ohs14eruma}.
The period of stage A superhumps from the times of
maxima ($E \le$ 13) is 0.0602(4)~d.  By using the data
between BJD 2457483.01 and 2457484.40, we have obtained
a period of 0.0601(1)~d with the phase dispersion minimization
(PDM; \cite{PDM}) method (figure \ref{fig:rzlmi2016shapdm}).
The errors are 1$\sigma$ estimated by the methods of
\citet{fer89error} and \citet{Pdot2}.  We consider that
the result by the PDM method is more reliable than
from superhump maxima since it gives a smaller error
and adopted this period.

\begin{figure}
  \begin{center}
    \FigureFile(80mm,110mm){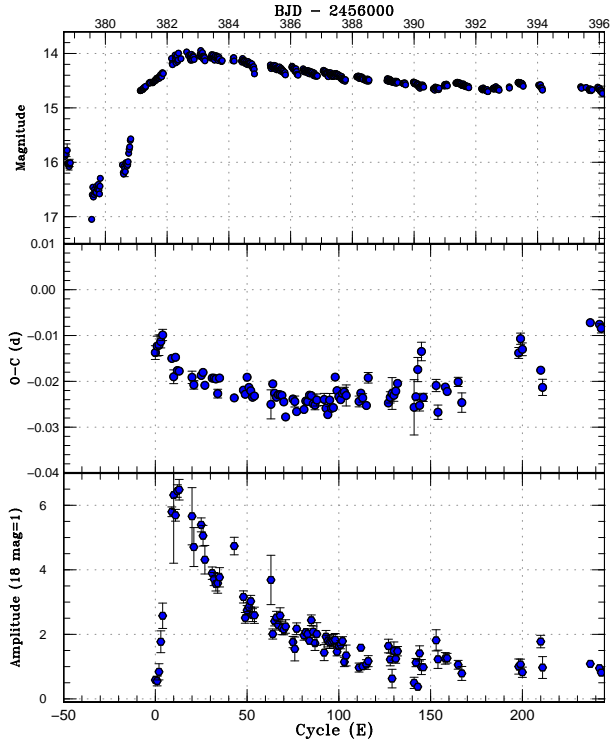}
  \end{center}
  \caption{$O-C$ diagram of superhumps in RZ LMi in 2013 April.
     (Upper:) Light curve.  The data were binned to 0.0023~d.
     (Middle:) $O-C$ diagram (filled circles).
     We used a period of 0.05945~d for calculating the $O-C$ residuals.
     (Lower:) Amplitudes of superhumps.  The scale is linear
     and the pulsed flux is shown in a unit corresponding
     to 18 mag = 1.
  }
  \label{fig:rzlmi2013humpall}
\end{figure}

\begin{figure}
  \begin{center}
    \FigureFile(80mm,70mm){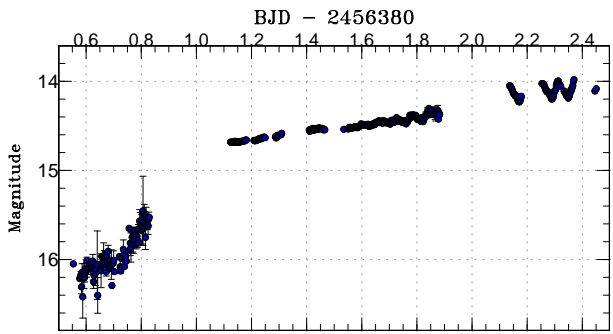}
  \end{center}
  \caption{Growing superhumps and stagnation (embedded
  precursor) at the start of the 2013 April superoutburst.
  The data were binned to 0.0023~d.
  Superhump grew between BJD 2456381.41 and 2456381.88.
  }
  \label{fig:rzlmi2013rise}
\end{figure}

\begin{figure}
  \begin{center}
    \FigureFile(80mm,110mm){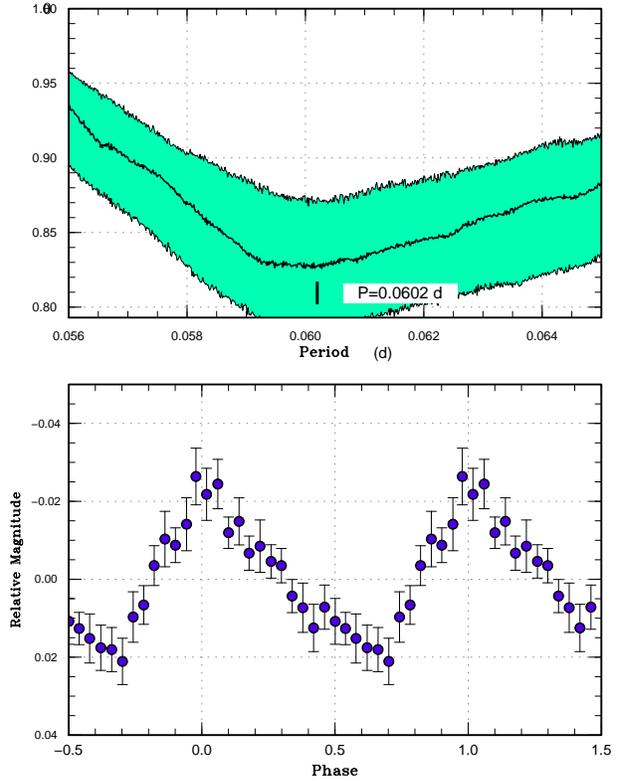}
  \end{center}
  \caption{Stage A superhumps in RZ LMi (2013 April).
     The data between BJD 2456381.41 and 2456381.88
     were used.
     (Upper): PDM analysis.
     (Lower): Phase-averaged profile.}
  \label{fig:rzlmi2013shapdm}
\end{figure}

   Although the second superoutburst was generally well
observed, the observations were unfortunately
relatively sparse around stage A.  The 2013 April
superoutburst of this object was relatively well observed
in the same phase, although the brightness peak was missed
and instead of a distinct precursor,
there was a stagnation in the rising phase, which was
likely an embedded precursor
(figures \ref{fig:rzlmi2013humpall}, \ref{fig:rzlmi2013rise};
see e-table for the full times of
superhump maxima).
During the observation in 2013, a continuous light curve
of growing superhumps (0$\le E \le$4) was well
recorded.  Using the data between BJD 2456381.41
and 2456381.88, we obtained a period of 0.0602(3)~d
with the PDM method (figure \ref{fig:rzlmi2013shapdm}).
Since the value is consistent with the 2016 one,
we adopted an averaged value of 0.0602(1)~d
as the period of stage A superhumps.

   The dynamical precession rate, $\omega_{\rm dyn}$
in the disk can be expressed by (see, \cite{hir90SHexcess}):
\begin{equation}
\label{equ:precession}
\omega_{\rm dyn}/\omega_{\rm orb} = Q(q) R(r),
\end{equation}
where $\omega_{\rm orb}$ and $r$ are the angular orbital frequency
and the dimensionless radius measured in units of the binary 
separation $a$.  The dependence on $q$ and $r$ are
\begin{equation}
\label{equ:qpart}
Q(q) = \frac{1}{2}\frac{q}{\sqrt{1+q}},
\end{equation}
and\footnote{
   There was a typographical error in the second line of equation (1)
in \citet{kat13qfromstageA}.  The correct formula is
$\frac{q}{\sqrt{1+q}} \bigl[\frac{1}{4}\sqrt{r} b_{3/2}^{(1)}\bigr]$.
The results (including tables) in \citet{kat13qfromstageA} used
the correct formula and the conclusion is unchanged.
We used the correct formula in equation (\ref{equ:rpart})
in this paper.  The same correction of the equation should
be applied to \citet{kat13j1222}, \citet{nak13j2112j2037}
and \citet{kat15wzsge}.
}
\begin{equation}
\label{equ:rpart}
R(r) = \frac{1}{2}\sqrt{r} b_{3/2}^{(1)}(r),
\end{equation}
where
$\frac{1}{2}b_{s/2}^{(j)}$ is the Laplace coefficient
\begin{equation}
\label{equ:laplace}
\frac{1}{2}b_{s/2}^{(j)}(r)=\frac{1}{2\pi}\int_0^{2\pi}\frac{\cos(j\phi)d\phi}
{(1+r^2-2r\cos\phi)^{s/2}}.
\end{equation}
This $\omega_{\rm dyn}/\omega_{\rm orb}$ is equal to
the fractional superhump excess in frequency:
$\epsilon^* \equiv 1-P_{\rm orb}/P_{\rm SH}$,
where $P_{\rm orb}$ and $P_{\rm SH}$ are the orbital
period and superhump period, respectively.
If $P_{\rm orb}$ is known, we can directly determine
$q$ from the observed $\epsilon^*$ of stage A superhumps
under the assumption that the period of stage A superhumps
reflects the purely dynamical precession rate
at the radius of the 3:1 resonance \citep{kat13qfromstageA}.

   Since the orbital period of RZ LMi is not known,
we cannot directly apply the method in \citet{kat13qfromstageA}
to dynamically determine $q$.
We can instead use the period of post-superoutburst
superhumps to constrain $q$ and the disk radius
as introduced in \citet{kat13j1222}:
\begin{equation}
\label{equ:epsstagea}
\epsilon^*({\rm stage A}) = Q(q) R(r_{\rm 3:1})
\end{equation}
and
\begin{equation}
\label{equ:epspost}
\epsilon^*({\rm post}) = Q(q) R(r_{\rm post}),
\end{equation}
where $r_{\rm 3:1}$ is the radius of the 3:1 resonance
\begin{equation}
\label{equ:radius31}
r_{3:1}=3^{(-2/3)}(1+q)^{-1/3},
\end{equation} 
$\epsilon^*({\rm post})$ and $r_{\rm post}$ are the fractional
superhump excess and disk radius immediately after
the outburst, respectively.  By solving equations
(\ref{equ:epsstagea}) and (\ref{equ:epspost}) simultaneously,
we can obtain the relation between $r_{\rm post}$ and $q$.
If we have knowledge about $r_{\rm post}$,
as determined in other systems in \citet{kat13qfromstageA},
we have a more stringent constraint.

\begin{figure}
  \begin{center}
    \FigureFile(80mm,110mm){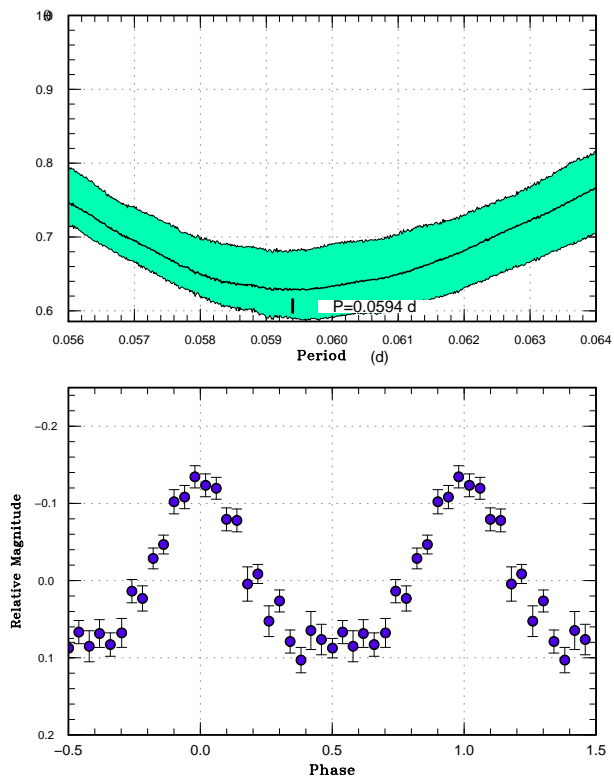}
  \end{center}
  \caption{Post-superoutburst superhumps in RZ LMi (2013 March 20).
     The data between BJD 2456371.5 and 2456371.9 were used.
     (Upper): PDM analysis.
     (Lower): Phase-averaged profile.}
  \label{fig:rzlmi2013shpostpdm}
\end{figure}

\begin{figure}
  \begin{center}
    \FigureFile(80mm,110mm){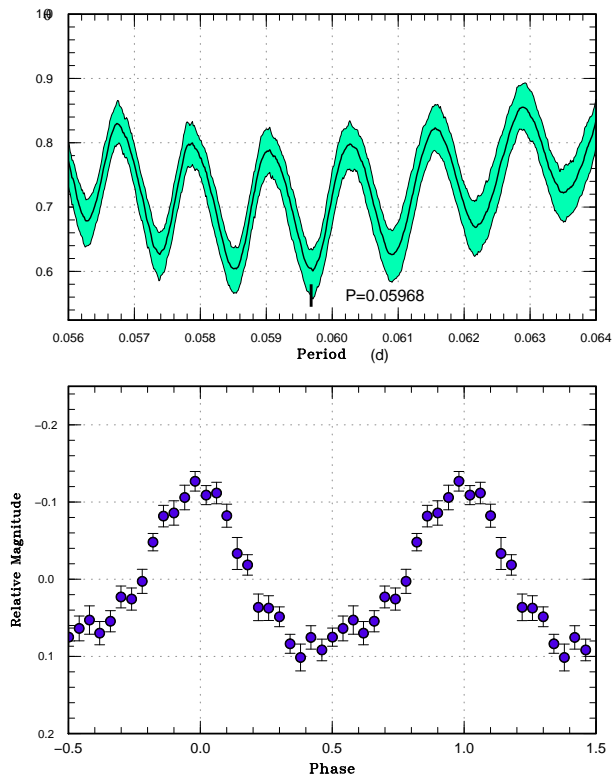}
  \end{center}
  \caption{Post-superoutburst superhumps in RZ LMi (2013 March 20 and 24).
     (Upper): PDM analysis.
     (Lower): Phase-averaged profile.}
  \label{fig:rzlmi2013shpostpdmall}
\end{figure}

   On 2013 March 20, the object was observed in
quiescence just following a superoutburst starting on
March 6.  The object displayed post-superoutburst
superhumps and there was a continuous run covering
six cycles.  A PDM analysis of this continuous run yielded
a period of 0.0594(2)~d (figure \ref{fig:rzlmi2013shpostpdm}).
Since this variation was also present after one more
normal outburst, and since the decaying amplitudes
of this variation exclude the possibility of
the orbital variation, we identified this period
as that of post-superoutburst superhumps.
We combined the quiescent data
on March 20 and 24 and obtained a period of
0.05969(2)~d (figure \ref{fig:rzlmi2013shpostpdmall},
assuming that the superhump phase and
period did not change during a normal outburst).
The shorter one-day alias ($\sim$0.0584~d) is too close
to the supposed orbital period (discussed later)
and this alias is unlikely.

\begin{figure}
  \begin{center}
    \FigureFile(80mm,70mm){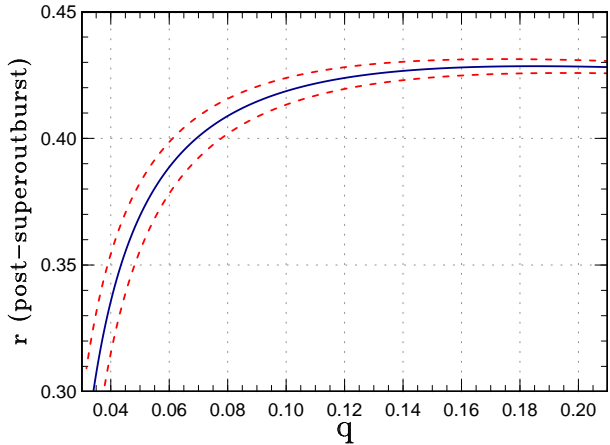}
  \end{center}
  \caption{Relation between $q$ and $r_{\rm post}$ using
  the periods of stage A superhumps and post-superoutburst
  superhumps.  Dashed curves represent the range of
  1$\sigma$ errors.}
  \label{fig:qrpost}
\end{figure}

   Using the periods of stage A superhumps and post-superoutburst
superhumps, the relation between $r_{\rm post}$ and $q$
is shown in figure \ref{fig:qrpost}.  The measurements
of $r_{\rm post}$ in SU UMa-type dwarf novae
using the same method are within the range of
0.30 and 0.38 \citep{kat13qfromstageA}.
The smaller values represent the values for
WZ Sge-type dwarf novae with multiple rebrightenings
(after such rebrightenings), and it is extremely unlikely
the case for RZ LMi.  If $r_{\rm post}$ is around 0.38,
$q$ is estimated to be 0.06(1).  It would be noteworthy
that \citet{osa95rzlmi} assumed $r_{\rm post}$=0.42
for this particular object.  If it is indeed the case,
$q$ needs to be as large as 0.10(2).  This consideration
suggests that the expectations by \citet{osa95rzlmi} and
\citet{hel01eruma} that RZ LMi has a very small $q$ and
the large disk radius immediately following a superoutburst
are not true at the same time: either $q$ is higher
or the disk radius is smaller.  This is the important
conclusion from the present observation.
Since stage A superhumps were not very ideally observed
in the present study, further observations are needed
to refine the result.

\subsection{Possible negative superhumps and orbital period}\label{sec:negSH}

\begin{figure}
  \begin{center}
    \FigureFile(80mm,95mm){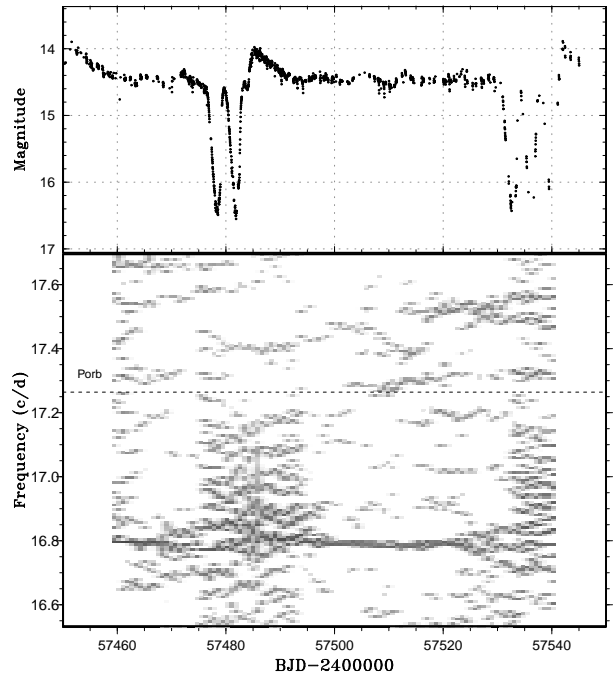}
  \end{center}
  \caption{Two-dimensional Lasso period analysis of RZ LMi
  (2016).
  (Upper:) Light curve.  The data were binned to 0.02~d.
  (Lower:) Lasso period analysis.  The strong persistent
  signal (disturbed during the starting phase of
  the superoutburst due to the non-sinusoidal profile and
  rapidly varying periods) around 16.8 c/d is superhumps.
  A weaker signal around 17.50--17.55 c/d between BJD
  2457510 and 2457530 is possible negative superhumps. 
  $\log \lambda=-6.4$ was used.  Porb represents
  the orbital period we suggest.
  The width of the sliding window and the time step used are
  18~d and 0.8~d, respectively.
  }
  \label{fig:rzlmi2016lasso}
\end{figure}

   We used least absolute shrinkage and selection operator (Lasso)
method (\cite{lasso}; \cite{kat12perlasso}), which has been 
proven to yield very sharp signals.  
We used the two-dimensional Lasso power spectra as introduced
in the analysis of the Kepler data such as in \citet{kat13j1924}; 
\citet{osa13v344lyrv1504cyg}.  These two-dimensional Lasso
power spectra have been proven to be very effective
in detecting signals in non-uniformly sampled ground-based data
(see e.g. \cite{ohs14eruma}) since Lasso-type period analysis
is less affected by aliasing than traditional Fourier-type
power spectra.  This characteristic has enabled detection
of negative superhumps in ground-based observations
(e.g. \cite{ohs14eruma}; \cite{Pdot6}).
The result for RZ LMi in 2016 is shown in figure
\ref{fig:rzlmi2016lasso}.  The strong persistent
signal around 16.8 cycle/day (c/d) is superhumps.
A weaker signal around 17.50--17.55 c/d between BJD
2457510 and 2457530 is possible negative superhumps. 
Since the second superoutburst lasted very long,
it may have been possible that negative superhumps
were excited during this long-lasting standstill-like
phase, which gives the condition almost the same
as in permanent superhumpers (see subsection \ref{sec:outburst}).
A PDM analysis of the data between BJD 2457510 and
2457530 yielded a period of 0.05710(1)~d.
The superhump period in this interval was 0.059555(4)~d.

   It has been widely accepted that absolute fractional
superhumps excesses ($\epsilon_+$ for positive superhumps,
$\epsilon_-$ for negative superhumps, where
$\epsilon \equiv P_{\rm SH}/P_{\rm orb}-1$) are tightly
correlated when both signals are simultaneously seen
in novalike CVs (e.g. \cite{pat97v603aql};
\cite{mon09diskprecession}).  The empirical relation is
$\epsilon_+ \simeq 2 |\epsilon_-|$.  If it is also
the case for RZ LMi, $P_{\rm orb}$ is expected to be
around 0.05792~d.  This period is labeled as Porb
in figure \ref{fig:rzlmi2016lasso}.
If this is indeed the orbital period, the consequence
is important (cf. subsection \ref{sec:stageA}).
The measured period of stage A superhumps [0.0602(1)~d]
gives $\epsilon^*$=0.038(2), which is equivalent to
$q$=0.105(5) (see table 1 in \cite{kat13qfromstageA}).
This value is higher than the typical
ones for WZ Sge-type dwarf novae, which have similar
$P_{\rm orb}$ as RZ LMi (\cite{kat13qfromstageA};
\cite{kat15wzsge}; \cite{Pdot8}).
This $q$ measurement invalidates the suggestion by
\citet{osa95rzlmi} and \citet{hel01eruma} that the unusual
properties of RZ LMi is a result of the very low $q$.
The $q$, however, is consistent with the relation
derived from the period of stage A and post-superoutburst
superhumps (subsection \ref{sec:stageA},
figure \ref{fig:qrpost}) assuming the large disk radius
at the end of a superoutburst.
If this interpretation is correct, the disk radius
at the end of a superoutburst is large as required by
\citet{osa95rzlmi}, but the origin of the large disk
(i.e. superoutbursts terminate earlier than in ordinary
SU UMa-type dwarf novae) cannot be attributed to
an exceptionally small $q$.  In recent years, some
SU UMa-type dwarf novae show early termination
of superoutbursts (such as V1006 Cyg, \cite{kat16v1006cyg}).
In \citet{kat16v1006cyg}, the termination may be
associated with the supposed appearance of stage C
superhumps (late-stage superhumps with a shorter constant
period).  The origin of stage C superhumps
is still poorly known, the reason of premature termination
of superoutbursts which may be related to the evolution
of stage C superhumps still needed to be clarified.

   There is supporting evidence for a large $q$:
the duration of stage A superhumps, which is considered
to reflect the growth time of the 3:1 resonance,
is very short (less than 1~d) in RZ LMi.  From the
theoretical standpoint, this growth time is expected
to be proportional to $1/q^2$ \citep{lub91SHa}, and
this relation has been confirmed in WZ Sge-type
dwarf novae \citep{kat15wzsge}.  As judged from
the rapid growth of superhumps in RZ LMi, it appears
extremely unlikely that RZ LMi has $q$ as small
as in WZ Sge-type dwarf novae.

\subsection{Evolutionary status}\label{sec:evol}

\begin{figure}
  \begin{center}
    \FigureFile(80mm,70mm){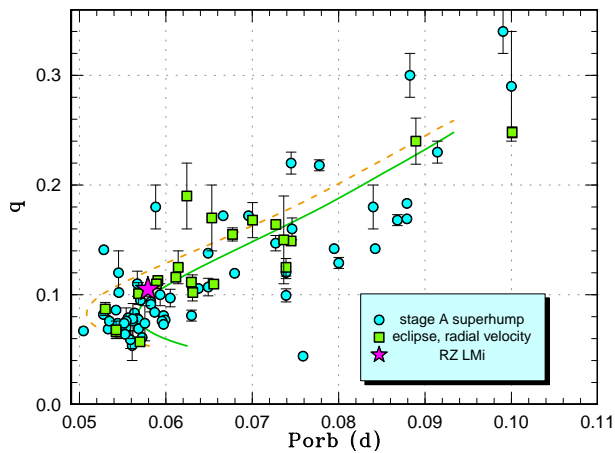}
  \end{center}
  \caption{Location of RZ LMi on the diagram of
  mass ratio versus orbital period.
  The dashed and solid curves represent the standard and optimal
  evolutionary tracks in \citet{kni11CVdonor}, respectively.
  The filled circles, filled squares, filled stars, filled diamonds
  represent $q$ values from a combination of the estimates
  from stage A superhumps published in four preceding
  sources (\cite{kat13qfromstageA}; \cite{nak13j2112j2037};
  \cite{Pdot5}; \cite{Pdot6}; \cite{Pdot7}; \citet{Pdot8}),
  known $q$ values from quiescent eclipses or 
  radial-velocity study (see \cite{kat13qfromstageA} for
  the data source).  In addition to the references listed in
  \citet{Pdot8}, we supplied the data for SDSS J143317.78$+$101123.3
  in \citet{her16j1433}.}
  \label{fig:qall5add}
\end{figure}

   If the $q$ derived from the possible negative superhumps
and suggested by the rapid growth of superhumps
(subsection \ref{sec:negSH}) is correct, RZ LMi
cannot be an object close to the period minimum
or a period bouncer --- the idea conjectured by
\citet{ole08rzlmi} that some ER UMa-type are
period bouncers.  The $q$ we suggested is similar to
or even larger than those of ordinary SU UMa-type dwarf novae
having similar $P_{\rm orb}$ (figure \ref{fig:qall5add}).

   We know at least another object GALEX J194419.33$+$491257.0
in the Kepler field which has a very short $P_{\rm orb}$
of 0.0528164(4)~d and very frequent outbursts
(normal outbursts with intervals of 4--10~d) \citep{kat14j1944}.
This object has an unusually high $q$=0.141(2) measured using
stage A superhumps.  These properties are somewhat similar
to those of RZ LMi.  \citet{kat14j1944} suggested
the possibility that GALEX J194419.33$+$491257.0
may be a CV with a stripped core evolved secondary which
are evolving toward AM CVn-type CVs.
Such a condition might be a fascinating possibility
to explain why RZ LMi has an exceptionally high $\dot{M}$
for its $P_{\rm orb}$.

\subsection{Secular variation of supercycle}\label{sec:secular}

   As discussed in subsection \ref{sec:outburst}, it is
likely that RZ LMi changed $\dot{M}$ by a factor
of $\sim$2 in the last two decades.
In recent years, a transition from the novalike (permanent
superhumper) state to the ER UMa-type state was discovered
in BK Lyn (\cite{pat13bklyn}; \cite{Pdot4}).
\citet{pat13bklyn} proposed, based on the potential
identification with an ancient classical nova in 101,
that ER UMa stars are transitional objects during the cooling
phase of post-eruption classical novae [the idea was not new
and it was already proposed in \citet{kat95eruma}
and \citet{osa95eruma}].
Following this interpretation, \citet{otu13suumacycle}
studied ER UMa-type objects and found a secular increase
of the supercycle in most of typical ER UMa-type objects.
RZ LMi was included, and \citet{otu13suumacycle} gave
$\dot{P}$ of the supercycle of (5.0$\pm$1.9)$\times 10^{-4}$
in 18 yr.  We should note that \citet{zem13eruma} also
studied variation of supercycles in ER UMa and reported
that supercycles vary in a range of 43.6--59.2~d
with shorter time-scales of 300--1900~d.  A secular increase
of the supercycle was also statistically meaningful
\citep{zem13eruma}.

   The conclusion of \citet{otu13suumacycle}, however,
was only based on the variation of supercycles, and
\citet{otu13suumacycle} disregarded the possibility
that the supercycle can also increase if $\dot{M}$
increases toward $\dot{M}_{\rm crit}$ since
there is a minimum in the supercycle as $\dot{M}$
increases (cf. the right branch of figure 1 in
\cite{osa95eruma}; see also subsection
\ref{sec:outburst}).  Since it is apparently the case for
RZ LMi, we studied secular variation of supercycles in
RZ LMi.  We summarized the result in
table \ref{tab:rzlmisupercycle} [the data in \citet{Pdot4}
also used AAVSO observations].  When raw data are not
available, we measured the fraction of superoutburst
and duty cycle by eye from the figures in the papers.
The data for 1986--1989 (Shugarov) were too sparse to
determine the supercycle and only the duty cycle was
estimated.  The data for 1987--1988 (Kato) were visual
observations.  The supercycle was determined from
seven well-defined bright outbursts.  The duty cycle
was probably underestimated due to the insufficient
detection limit of visual observations.
For AAVSO observations, the fraction of superoutburst
could be determined only to 0.05 since most of the data
were randomly sampled snapshot observations.

\begin{figure}
  \begin{center}
    \FigureFile(80mm,70mm){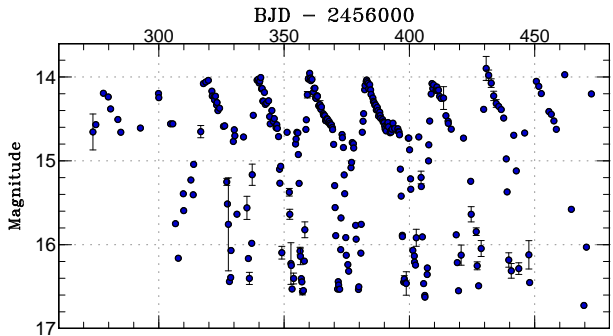}
  \end{center}
  \caption{Overall light curve of RZ LMi in 2013.
  The data were binned to 0.03~d.
  }
  \label{fig:rzlmi2013all}
\end{figure}

\begin{figure}
  \begin{center}
    \FigureFile(80mm,100mm){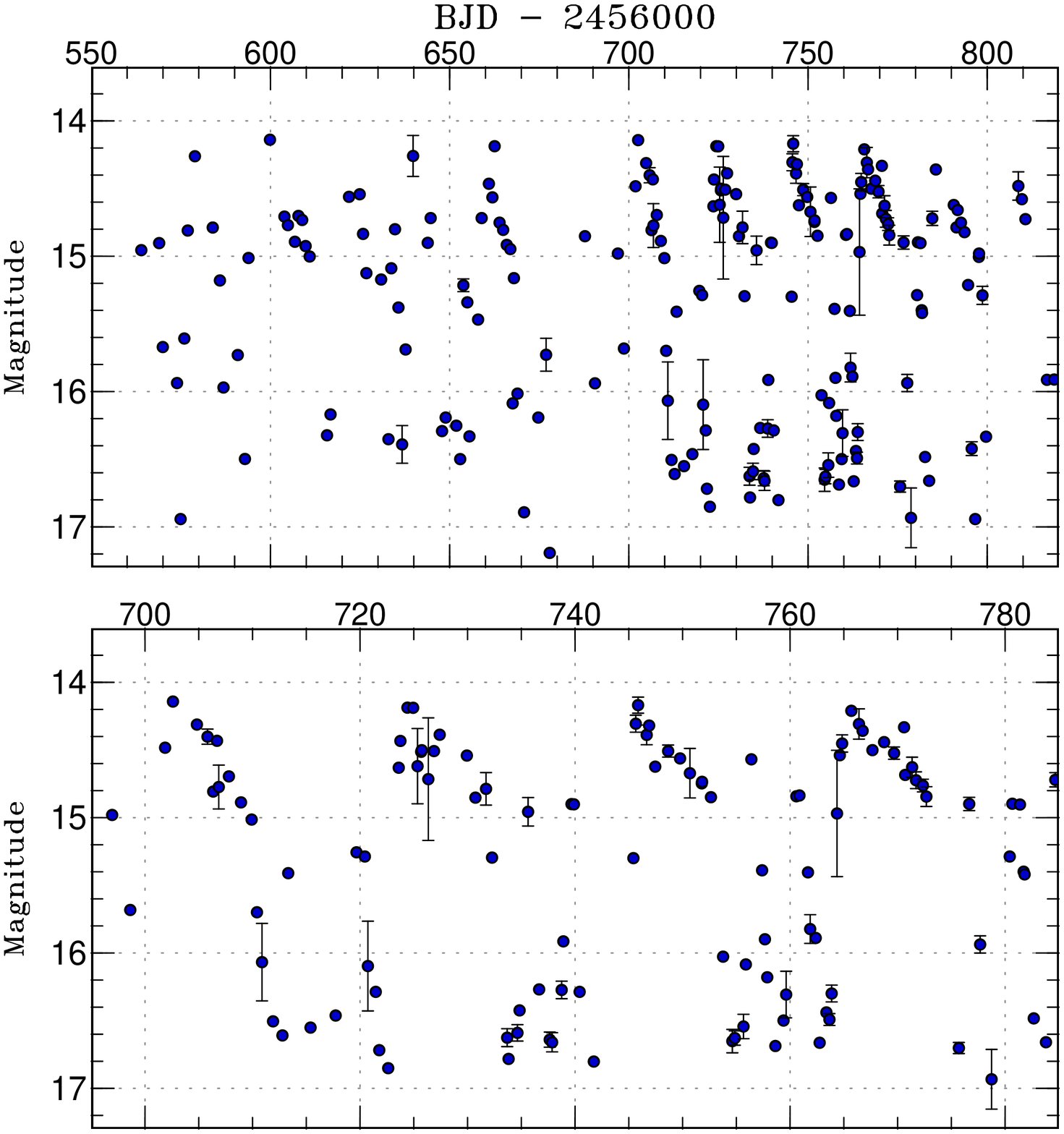}
  \end{center}
  \caption{Overall light curve of RZ LMi in 2013--2014.
  The data were binned to 0.03~d.
  }
  \label{fig:rzlmi2014all}
\end{figure}

   It has became apparent that the supercycle was not stable
as had been supposed in \citet{rob95eruma} and \citet{ole08rzlmi}.
The supercycle at the time of \citet{rob95eruma} probably
reached the historical minimum.  It was likely the supercycle
once lengthened as long as to $\sim$24~d before
2013, but it again returned to 19--20~d.  This behavior
probably gave the impression that the supercycle
is globally constant if seen in long time scales,
such as several years.
The situation changed in 2016 when the supercycle
strongly increased.  This increase was associated
with the increase of the fraction of superoutburst
and the duty cycle, indicating that $\dot{M}$ increased
despite the lengthening of the supercycle (on the contrary
to the conclusion by \cite{otu13suumacycle}).
Although the situation before 2016 was less clear,
the changing supercycle suggests fluctuating $\dot{M}$
within time scales of a year or two.
The outburst pattern was generally regular in the second
best observed season in 2013 (figure \ref{fig:rzlmi2013all}).
The duration of the superoutburst was appreciably longer
between BJD 2456380 and 2456400 (17~d, in contrast to
12--13~d in other superoutbursts in 2013;
a mean value of supercycle in 2013 is given in 
table \ref{tab:rzlmisupercycle}).
It was likely that $\dot{M}$ temporarily increased
also in 2013.  In the 2013--2014 season, the durations
of superoutbursts again decreased to shorter than 12~d
(figure \ref{fig:rzlmi2014all}) while supercycle lengths
remained short ($\sim$21~d).

\begin{table*}
\caption{Supercycles of RZ LMi}\label{tab:rzlmisupercycle}
\begin{center}
\begin{tabular}{ccccccc}
\hline
Year & JD range\commenta & supercycle (d) & fraction of superoutburst & duty cycle\commentb & nights & source \\
\hline
1984       & 45699--45851 & 19.6(1) & 0.4   & 0.53 & 30 & this work (Shugarov) \\
1984--1985 & 46019--46230 & 19.5--21.5 & 0.35: & 0.39 & 31 & this work (Shugarov) \\
1986--1989 & 46468--47682 & -- & -- & 0.6 & 7 & this work (Shugarov) \\
1987--1988 & 47151--47318 & 22.5(9) & -- & 0.39 & 52 & this work (Kato) \\
1992--1995 & 48896--49723 & 18.87   & 0.4 & 0.5 & -- & \citet{rob95eruma} \\
2004--2005 & 53027--53519 & 19.07(4) & 0.4 & 0.4 & -- & \citet{ole08rzlmi} \\
2005--2006 & 53644--53884 & 19.7(1) & 0.45 & 0.58 & 89 & AAVSO \\
2006--2007 & 54041--54254 & 20.75(4) & 0.40 & 0.55 & 113 & AAVSO \\
2007--2008 & 54374--54626 & 21.3(1) & 0.45 & 0.55 & 100 & AAVSO \\
2008--2009 & 54749--54988 & 21.7(1) & 0.40 & 0.31 & 96 & AAVSO \\
2009--2010 & 55146--55334 & 24.3(1) & 0.35 & 0.52 & 58 & AAVSO \\
2012       & 55985--56070 & 21.61(2) & 0.55 & 0.46 & 76 & \citet{Pdot4} \\
2012--2013 & 56273--56473 & 22.83(1) & 0.50 & 0.62 & 149 & this work \\
2013--2014 & 56563--56819 & 20.8(1) & 0.40 & 0.48 & 151 & this work (Neustroev), AAVSO \\
2014--2015 & 56930--57178 & 19.9(1) & 0.45 & 0.55 & 96 & AAVSO \\
2016       & 57416--57451 & 35 & 0.60 & 0.50 & 28 & this work \\
2016       & 57451--57483 & 32 & 0.81 & 0.74 & 31 & this work \\
2016       & 57483--57543 & 60 & 0.80 & 0.83 & 54 & this work \\
\hline
  \multicolumn{5}{l}{\commenta JD$-$2400000.} \\
  \multicolumn{5}{l}{\commentb Object brighter than 15.0 mag.} \\
\end{tabular}
\end{center}
\end{table*}

   As far as RZ LMi is concerned, $\dot{M}$ is not secularly
{\rm decreasing} as in the scenario by \citet{pat13bklyn}.
Among other ER UMa-type objects, BK Lyn returned back
to its original novalike state in late 2013 and
the ER UMa-type state was only transiently present
\citep{Pdot4}.  As discussed in \citet{Pdot4},
the hypothetical cooling sequence from novalike objects --
ER UMa-type dwarf novae -- ordinary SU UMa-type dwarf novae
after nova eruptions as proposed by \citet{pat13bklyn}
is not very consistent with observational statistics,
since such a cooling sequence would predict a large
number of intermediate (having much more slowly cooling
white dwarfs) objects between ER UMa-type dwarf novae 
and ordinary SU UMa-type dwarf novae, while observations
could not confirm a large number of such objects.
As seen in ER UMa, BK Lyn and RZ LMi, the $\dot{M}$
variations look more irregular with time-scales of
several years and it looks like that the majority of
variations in outburst activities in these systems 
does not reflect the secular $\dot{M}$ variation
as proposed by \citet{pat13bklyn}.
As this paper has shown, the high activity of RZ LMi
may be a result of a rare evolutionary condition
with a relatively massive secondary.  In such a case,
we may not require a nova eruption to produce
an object like RZ LMi.

\section{Summary}\label{sec:summary}

   We observed RZ LMi, which is renowned for the extremely
($\sim$19~d) short supercycle and is a member of
a small, unusual class of cataclysmic variable
called ER UMa-type dwarf novae, in 2013 and 2016.
In 2016, the supercycles of this object substantially
lengthened in comparison to the previous measurements
to 35, 32, 60~d for three consecutive
superoutbursts.  The durations of superoutbursts
also lengthened and they composed 60--81\% of
the supercycle.  Such long durations of superoutbursts
have never been observed in this object, and we consider
that the object virtually experienced a transition
to the novalike state (permanent superhumper).
This observed behavior extremely well reproduced
the prediction of the thermal-tidal instability model
by \citet{osa95eruma}.  We estimated that in 2016,
the mass-transfer rate of RZ LMi reached 97--99\%
of the thermal stability limit, and that it was about
two times larger than the value in the past.

   We detected a precursor in the 2016 superoutburst
and detected growing (stage A) superhumps in 2016
and in 2013.  We estimated their period to be
0.0602(1)~d.  This makes the second case in ER UMa-type
dwarf novae in which growing superhumps were confidently
recorded during the sequence of the precursor and
the main superoutburst.
We also detected post-superoutburst
superhumps immediately after a superoutburst in 2013
with a period of 0.05969(2)~d.
Since both stage A superhumps and post-superoutburst
superhumps can be considered to reflect the dynamical
precession rates, we can derive the relation between
the mass ratio and the radius of the post-superoutburst.
The result suggests that the mass ratio is not
as small as in WZ Sge-type dwarf novae, having orbital
periods similar to RZ LMi.

   By using least absolute shrinkage and selection
operator (Lasso) two-dimensional power spectra,
we detected possible negative superhumps with
a period of 0.05710(1)~d.  By combination with
the period of ordinary superhumps, we estimated
the orbital period of 0.05792~d.  The period of
stage A superhumps, combined with this orbital period,
suggests a mass ratio of 0.105(5).  This relatively
large mass ratio is consistent with the rapid growth
of superhumps.  This mass ratio is even above
ordinary SU UMa-type dwarf novae, and it is also 
possible that the exceptionally high mass-transfer rate 
in RZ LMi may be a result of a stripped core evolved
secondary in a system evolving toward an AM CVn-type object.

   An analysis of historical records of supercycles
in this system suggests that the variation of
the outburst activity is sporadic with time-scales
of years and it does not seem to reflect
the secular variation caused by evolution.

\section*{Acknowledgments}

This work was supported by the Grant-in-Aid
``Initiative for High-Dimensional Data-Driven Science through Deepening
of Sparse Modeling'' (25120007) 
from the Ministry of Education, Culture, Sports, 
Science and Technology (MEXT) of Japan.
CCN thanks the funding from National Science Council of Taiwan 
under the contract NSC101-2112-M-008-017-MY3.
This work also was partially supported by
RFBR grant 15-02-06178 (Crimean team, Shugarov and Katysheva) and
Grant VEGA No. 2/0002/13 (Shugarov, Chochol, Seker\'a\v{s}).
The authors are grateful to observers of VSNET Collaboration and
VSOLJ observers who supplied vital data.
We acknowledge with thanks the variable star
observations from the AAVSO International Database contributed by
observers worldwide and used in this research.
This research has made use of the SIMBAD database,
operated at CDS, Strasbourg, France.

\section*{Supporting information}

Additional supporting information can be found in the online version
of this article.
Supplementary data is available at PASJ Journal online.


\begin{thebibliography}{}

\bibitem[{Abrahamian}, {Mickaelian}(1993)]{abr93FBS}
  {Abrahamian}, H.~V., \& {Mickaelian}, A.~M.\ 1993, Astrofizika, 36, 109

\bibitem[{Cleveland}(1979)]{LOWESS}
  {Cleveland}, W.~S.\ 1979, J. Amer. Statist. Assoc., 74, 829

\bibitem[Fernie(1989)]{fer89error}
  Fernie, J.~D.\ 1989, PASP, 101, 225

\bibitem[Green et~al.(1982)]{gre82PGsurveyCV}
  Green, R.~F., Ferguson, D.~H., Liebert, J., \& Schmidt, M.\ 1982, PASP, 94,
  560

\bibitem[Hellier(2001)]{hel01eruma}
  Hellier, C.\ 2001, PASP, 113, 469

\bibitem[{Hern{\'a}ndez Santisteban} et~al.(2016)]{her16j1433}
  {Hern{\'a}ndez Santisteban}, J.~V., {et~al.}\ 2016, Nature, 533, 366

\bibitem[{Hirose}, {Osaki}(1990)]{hir90SHexcess}
  {Hirose}, M., \& {Osaki}, Y.\ 1990, PASJ, 42, 135

\bibitem[Iida(1994)]{iid94eruma}
  Iida, M.\ 1994, VSOLJ\ Variable\ Star\ Bull., 19, 2

\bibitem[{Kato}(2015)]{kat15wzsge}
  {Kato}, T.\ 2015, PASJ, 67, 108

\bibitem[{Kato} et~al.(2014a)]{Pdot6}
  {Kato}, T., {et~al.}\ 2014a, PASJ, 66, 90

\bibitem[{Kato} et~al.(2015)]{Pdot7}
  {Kato}, T., {et~al.}\ 2015, PASJ, 67, 105

\bibitem[{Kato} et~al.(2013a)]{Pdot4}
  {Kato}, T., {et~al.}\ 2013a, PASJ, 65, 23

\bibitem[{Kato} et~al.(2014b)]{Pdot5}
  {Kato}, T., {et~al.}\ 2014b, PASJ, 66, 30

\bibitem[{Kato} et~al.(2016a)]{Pdot8}
  {Kato}, T., {et~al.}\ 2016a, PASJ, in press (arXiv/1605.06221)

\bibitem[{Kato} et~al.(2009)]{Pdot}
  {Kato}, T., {et~al.}\ 2009, PASJ, 61, S395

\bibitem[{Kato}, {Kunjaya}(1995)]{kat95eruma}
  {Kato}, T., \& {Kunjaya}, C.\ 1995, PASJ, 47, 163

\bibitem[{Kato}, {Maehara}(2013)]{kat13j1924}
  {Kato}, T., \& {Maehara}, H.\ 2013, PASJ, 65, 76

\bibitem[{Kato} et~al.(2010)]{Pdot2}
  {Kato}, T., {et~al.}\ 2010, PASJ, 62, 1525

\bibitem[{Kato} et~al.(2013b)]{kat13j1222}
  {Kato}, T., {Monard}, B., {Hambsch}, F.-J., {Kiyota}, S., \& {Maehara}, H.\
  2013b, PASJ, 65, L11

\bibitem[Kato et~al.(1999)]{kat99erumareview}
  Kato, T., Nogami, D., Baba, H., Masuda, S., Matsumoto, K., \& Kunjaya, C.\
  1999, in Disk Instabilities in Close Binary Systems, ed. S. Mineshige, \&
  J.~C. Wheeler (Tokyo: Universal Academy Press), p.~45

\bibitem[{Kato}, {Osaki}(2013)]{kat13qfromstageA}
  {Kato}, T., \& {Osaki}, Y.\ 2013, PASJ, 65, 115

\bibitem[{Kato}, {Osaki}(2014)]{kat14j1944}
  {Kato}, T., \& {Osaki}, Y.\ 2014, PASJ, 66, L5

\bibitem[{Kato} et~al.(2016b)]{kat16v1006cyg}
  {Kato}, T., {et~al.}\ 2016b, PASJ, 68, L4

\bibitem[{Kato}, {Uemura}(2012)]{kat12perlasso}
  {Kato}, T., \& {Uemura}, M.\ 2012, PASJ, 64, 122

\bibitem[Kato et~al.(2004)]{VSNET}
  Kato, T., Uemura, M., Ishioka, R., Nogami, D., Kunjaya, C., Baba, H., \&
  Yamaoka, H.\ 2004, PASJ, 56, S1

\bibitem[Kholopov et~al.(1985)]{NameList67}
  Kholopov, P.~N., Samus, N.~N., Kazarovets, E.~V., \& Perova, N.~B.\ 1985,
  IBVS, 2681

\bibitem[{Knigge} et~al.(2011)]{kni11CVdonor}
  {Knigge}, C., {Baraffe}, I., \& {Patterson}, J.\ 2011, ApJS, 194, 28

\bibitem[Kondo et~al.(1984)]{kon84KUV2}
  Kondo, M., Noguchi, T., \& Maehara, H.\ 1984, Tokyo\ Astron.\ Obs.\ Annals,\
  Sec.\ Ser., 20, 130

\bibitem[Lipovetskii, Stepanyan(1981)]{lip81FBSCV}
  Lipovetskii, V.~A., \& Stepanyan, J.~A.\ 1981, Astrofizika, 17, 573

\bibitem[{Lubow}(1991)]{lub91SHa}
  {Lubow}, S.~H.\ 1991, ApJ, 381, 259

\bibitem[Misselt, Shafter(1995)]{mis95PGCV}
  Misselt, K.~A., \& Shafter, A.~W.\ 1995, AJ, 109, 1757

\bibitem[{Montgomery}(2009)]{mon09diskprecession}
  {Montgomery}, M.~M.\ 2009, ApJ, 705, 603

\bibitem[{Nakata} et~al.(2013)]{nak13j2112j2037}
  {Nakata}, C., {et~al.}\ 2013, PASJ, 65, 117

\bibitem[Nogami et~al.(1995)]{nog95rzlmi}
  Nogami, D., Kato, T., Masuda, S., Hirata, R., Matsumoto, K., Tanabe, K., \&
  Yokoo, T.\ 1995, PASJ, 47, 897

\bibitem[{Ohshima} et~al.(2014)]{ohs14eruma}
  {Ohshima}, T., {et~al.}\ 2014, PASJ, 66, 67

\bibitem[{Olech} et~al.(2008)]{ole08rzlmi}
  {Olech}, A., {Wisniewski}, M., {Zloczewski}, K., {Cook}, L.~M., {Mularczyk},
  K., \& {Kedzierski}, P.\ 2008, Acta\ Astron., 58, 131

\bibitem[{Osaki}(1989)]{osa89suuma}
  {Osaki}, Y.\ 1989, PASJ, 41, 1005

\bibitem[{Osaki}(1995a)]{osa95eruma}
  {Osaki}, Y.\ 1995a, PASJ, 47, L11

\bibitem[{Osaki}(1995b)]{osa95rzlmi}
  {Osaki}, Y.\ 1995b, PASJ, 47, L25

\bibitem[{Osaki}, {Kato}(2013a)]{osa13v1504cygKepler}
  {Osaki}, Y., \& {Kato}, T.\ 2013a, PASJ, 65, 50

\bibitem[{Osaki}, {Kato}(2013b)]{osa13v344lyrv1504cyg}
  {Osaki}, Y., \& {Kato}, T.\ 2013b, PASJ, 65, 95

\bibitem[{Otulakowska-Hypka}, {Olech}(2013)]{otu13suumacycle}
  {Otulakowska-Hypka}, M., \& {Olech}, A.\ 2013, MNRAS, 433, 1338

\bibitem[Patterson(1999)]{pat99SH}
  Patterson, J.\ 1999, in Disk Instabilities in Close Binary Systems, ed. S.
  Mineshige, \& J.~C. Wheeler (Tokyo: Universal Academy Press), p.~61

\bibitem[Patterson et~al.(1997)]{pat97v603aql}
  Patterson, J., Kemp, J., Saad, J., Skillman, D.~R., Harvey, D., Fried, R.,
  Thorstensen, J.~R., \& Ashley, R.\ 1997, PASP, 109, 468

\bibitem[{Patterson} et~al.(2013)]{pat13bklyn}
  {Patterson}, J., {et~al.}\ 2013, MNRAS, 434, 1902

\bibitem[Pikalova, Shugarov(1995)]{pik95rzlmi}
  Pikalova, O.~D., \& Shugarov, S.~{\relax Yu.}\ 1995, in Cataclysmic
  Variables, ed. A. Bianchini, M. della Valle, \& M. Orio (Dordrecht: Kluwer
  Academic Publishers), p.~173

\bibitem[{Robertson} et~al.(1994)]{rob94rzlmi}
  {Robertson}, J.~W., {Honeycutt}, R.~K., \& {Turner}, G.~W.\ 1994, in ASP\
  Conf.\ Ser.\ 56, Interacting Binary Stars, ed. A.~W. {Shafter} Vol.~56(.
San Francisco: ASP), p.~298

\bibitem[Robertson et~al.(1995)]{rob95eruma}
  Robertson, J.~W., Honeycutt, R.~K., \& Turner, G.~W.\ 1995, PASP, 107, 443

\bibitem[{Skillman}, {Patterson}(1993)]{ski93bklyn}
  {Skillman}, D.~R., \& {Patterson}, J.\ 1993, ApJ, 417, 298

\bibitem[Stellingwerf(1978)]{PDM}
  Stellingwerf, R.~F.\ 1978, ApJ, 224, 953

\bibitem[{Tibshirani}(1996)]{lasso}
  {Tibshirani}, R.\ 1996, J. R. Statistical Soc. Ser. B, 58, 267

\bibitem[Warner(1995)]{war95book}
  Warner, B.\ 1995, Cataclysmic Variable Stars (Cambridge: Cambridge University
  Press)

\bibitem[{Zemko} et~al.(2013)]{zem13eruma}
  {Zemko}, P., {Kato}, T., \& {Shugarov}, S.\ 2013, PASJ, 65, 54

\end{thebibliography}
\end{document}